\documentclass[12pt]{article}
\begin{document}

\title{\bf Timelike and Spacelike Matter Inheritance Vectors in Specific Forms
of Energy-Momentum Tensor}

\author{M. Sharif \thanks{e-mail: msharif@math.pu.edu.pk} and Umber
Sheikh
\\ Department of Mathematics, University of the Punjab,\\ Quaid-e-Azam
Campus Lahore-54590, PAKISTAN.}

\date{}

\maketitle

\begin{abstract}
This paper is devoted to the investigation of the consequences of
timelike and spacelike matter inheritance vectors in specific
forms of energy-momentum tensor, i.e., for string cosmology
(string cloud and string fluid) and perfect fluid. Necessary and
sufficient conditions are developed for a spacetime with string
cosmology and perfect fluid to admit a timelike matter inheritance
vector, parallel to $u^a$ and spacelike matter inheritance vector,
parallel to $x^a$. We compare the outcome with the conditions of
conformal Killing vectors. This comparison provides us the
conditions for the existence of matter inheritance vector when it
is also a conformal Killing vector. Finally, we discuss these
results for the existence of matter inheritance vector in the
special cases of the above mentioned spacetimes.
\end{abstract}


\date{}

\section{Introduction}

General Relativity (GR) is the field theory of gravity which is
highly non-linear. The non-linearity of GR makes the field
equations non-linear. Symmetries yield physical restrictions to
the gravitational field which give not only the simplicity but
also provide special physical effects in the field. GR provides a
rich arena to use symmetries in order to understand the natural
relation between geometry and matter furnished by Einstein field
equations (EFEs).

Symmetry inheritance of a kinematical or dynamical quantity is
defined [1] mathematically as
\begin{equation}
\pounds_{\xi} A = 2 \alpha A,
\end{equation}
where $\pounds_{\xi}$ is Lie derivative along ${\xi}$,
${\xi}(x^a)$ is the inheritance vector, $\alpha(x^a)$ is a scalar
function, and $A$ is any of the following: $g_{ab}$ (metric
tensor), $R^a_{bcd}$ (Riemann tensor), $R_{ab}$ (Ricci tensor),
$T_{ab}$ (matter tensor) or geometric objects constructed by them.
One can find all well-known inheritance symmetries by requiring
the particular forms of the quantity $A$. For example, if we take
$A = g_{ab}$, the above equation defines an inheritance symmetry
defined by Conformal Killing Vector (CKV). If $A = R^a_{bcd}$,
this defines Riemann inheritance symmetry or Curvature inheritance
symmetry. If $A = R_{ab}$, this is Ricci inheritance symmetry and
for $A = T_{ab}$, the equation defines the matter inheritance
symmetry. In the case of inheritance symmetry of CKVs, the
function $\alpha(x^a)$ is called the conformal factor and in the
case of other inheritance symmetries, it is called the inheritance
factor. When $\alpha = 0$, all the inheritance cases reduce to the
cases of collineations.

The matter inheritance symmetry (MIV) is defined by
\begin{equation}
\pounds_{\xi} T_{ab} = 2 \alpha T_{ab},
\end{equation}
which can be expressed in component form as
\begin{equation}
T_{ab;c} \xi^c + T_{ac} \xi^c_{;b} + T_{bc} \xi^c_{;a} = 2 \alpha
T_{ab}.
\end{equation}
The study of MIV is important as it helps in studying the
invariance properties of a given geometrical object namely the
Einstein tensor. This tensor plays an important role in the theory
of GR, since it is related via EFEs to the material content of
spacetime (represented by the matter tensor). Also, the symmetries
of energy-momentum tensor provide conservation laws on matter
field. They also help us to find out how the physical fields
occurring in a certain region of spacetime reflect the symmetries
of the metric.

A recent literature on inheritance symmetries in different
spacetimes has attracted many people. Herrera and Ponce [2] have
discussed CKVs in perfect and anisotropic fluids. Maartens
\textit{et al.} [3] have made a study of special conformal Killing
vectors (SCKVs) in anisotropic fluids.  Coley and Tupper [4] have
considered spacetimes admitting SCKV and symmetry inheritance.
Carot \textit{et al.} [5] have discussed spacetimes with CKVs.
Duggal [1,6] has studied curvature inheritance symmetry and
timelike Ricci inheritance symmetry in fluid spacetimes.

Greenberg [7] was the first to introduce the theory of spacelike
congruences in GR. It was further developed with applications by
Mason and Tsamparlis [8], who considered spacelike CKVs in
spacelike congruences. Yavuz and Yilmaz [9] considered inheriting
CKVs and SCKVs in string cosmology (string cloud and string
fluid). They also discussed solutions of string cosmology in
static spherical symmetric spacetime via CKVs. Yilmaz \textit{et
al.} [10] and Baysal \textit{et al.} [11] worked on the curvature
inheritance symmetry and conformal collineations respectively in
string cosmology. In his paper, Yilmaz [12] has considered
timelike and spacelike Ricci collineations in string cloud. Baysal
and Yilmaz [13] extended this work to spacelike Ricci inheritance
vectors in string cosmology. In another paper [14], the same
authors studied timelike and spacelike Ricci collineations in the
model of string fluid.

Zeldovich [15] argued that the study of string cloud and string
fluid models could give rise to density perturbations leading to
the formation of galaxies. Kibble [16] shows the consistency in
the existence of a large scale network of strings in the early
universe and the today's observations of the universe. According
to them the grand unification theories can also explain the
presence of strings. Thus, it would be worth interesting to
investigate the symmetry features of strings. The energy-momentum
tensor associated with a perfect fluid has been widely studied in
GR as a source of gravitational field, mainly to describe models
of stars, galaxies and universes [17]. Also, it has been used to
solve the EFEs using different prescriptions. Here the symmetry
feature of the perfect fluid would be used.

In this paper, we extend the work for timelike MIVs and spacelike
MIVs (SpMIVs) using string cosmology and perfect fluid. We shall
also discuss the conditions for a MIV to be a CKV. The paper has
been organised as follows. In section 2, we shall review some
general results of timelike MIVs as well as SpMIVs to be used in
the next sections. In sections 3 and 4, we shall solve the MIV
equations and find out the necessary and sufficient conditions for
timelike MIVs and SpMIVs in different fluid spacetimes to admit
them. Also, we shall give the necessary and sufficient conditions
for both timelike MIV and SpMIV when it is a CKV. In the last
section, we shall conclude and discuss the results obtained.

\section{Some General Results}

This section is devoted to discuss some general results about
timelike MIVs as well as SpMIVs which will be used in later
sections. Before discussing these results, we give the specific
forms of the matter tensor to be used in this paper.

The energy-momentum tensor for a string cloud can be written as
[18]
\begin{equation}
T_{ab}=\rho u_au_b-\lambda x_ax_b,
\end{equation}
where $\rho$ is the rest energy for the cloud of strings with
particles attached to them and $\lambda$ is the string tensor
density and are related by $\rho=\rho_p+\lambda.$ Here $\rho_p$ is
particle energy density. This energy-momentum tensor represents a
model of massive strings. Each massive string is formed by a
geometric string with particles attached along its extension. This
is the simplest model where we have particle and strings together.

The energy-momentum tensor for a string fluid [20] is
\begin{equation}
T_{ab}=\rho_s(u_au_b-x_ax_b)+qH_{ab},
\end{equation}
where $\rho_s$ is the string density, $q$ is the string tension
and also pressure. $H_{ab}$ is screen projection operator and will
be explained later. The energy-momentum tensor for a perfect fluid
[17] is
\begin{equation}
T_{ab}=(\rho+p)u_au_b+p g_{ab},
\end{equation}
where $\rho$ is the energy density and p is the pressure of fluid.

\subsection{Timelike Matter Inheritance Vectors}

Let $\xi^a=\xi u^a$, where $u^a$ is a unit timelike four-velocity
vector orthogonal to the four-vector $x^a$ satisfying the
following properties:
\begin{equation}
x^a x_a=1,\quad\ u^a u_a=-1 \quad\ u^a x_a=0.
\end{equation}
The vector $u_{a;b}$ can be decomposed into its antisymmetric
part, symmetric trace free part and trace as follows [20]
\begin{equation}
u_{a;b}= \omega_{ab}+\sigma_{ab}+\frac{\theta}{3}h_{ab}-\dot{u}_a
u_b,
\end{equation}
where $\dot{u}_a$ is the acceleration, $\omega_{ab}$ is rotation
(rotational velocity), $\sigma_{ab}$ is shear (shear velocity) and
$\theta$ is the expansion (expansion velocity). These quantities
are mathematically defined as
\begin{eqnarray}
\dot{u}_a &=& u_{a;n}u^n=\frac{Du_a}{D\tau},\\
\omega_{ab} &=& u_{[a;b]}+\dot{u}_{[a}u_{b]},\\
\sigma_{ab} &=& u_{(a;b)}+\dot{u}_{(a}u_{b)}
-\frac{\theta}{3}h_{ab},\\
\theta &=& u^a_{;a}.
\end{eqnarray}
The projection tensor, $h_{ab}=g_{ab}+u_au_b,$ has the following
properties:
\begin{equation}
h^{ab}x_a=x^b,\quad h^{ab}u_a=0,\quad h_c^a h_b^c=h_b^a,\quad
h_{ab}=h_{ba},\quad h_a^a=3.
\end{equation}

When $\xi^a=\xi u^a$, we can re-write the matter inheritance
symmetry as
\begin{equation}
\dot{T}_{ab} + u^cT_{c(a} ln\xi_{,b)} + T_{c(a} u^c_{;b)} =
2\alpha \xi^{-1} T_{ab}.
\end{equation}

The primary effect of a timelike CKV $\xi^a=\xi u^a$ is a
well-known equation given by $ \pounds_{\xi} g_{ab} = 2 \psi
g_{ab}$ which is equivalent to the following conditions [21]
(Proof is given in Appendix A)
\begin{eqnarray}
\dot{u}_a -(ln \xi)_{,a}&=& u_a(ln\xi)^.=\psi\xi^{-1}u_a ,\\
\theta&=& 3\psi\xi^{-1},\\
\sigma_{cd} &=&0.
\end{eqnarray}

\subsection{Spacelike Matter Inheritance Vectors}

Let $\xi^a=\xi x^a$, where $x^a$ is a unit spacelike vector
orthogonal to unit four-velocity vector $u^a$ satisfying the
following relations:
\begin{equation}
x^a x_a=1,\quad\ u^a u_a=-1 \quad\ u^a x_a=0.
\end{equation}
The vector $x_{a;b}$ can be decomposed with respect to $u^a$ and
$x^a$ in the following way [22]
\begin{equation}
x_{a;b}=A_{ab}+x_a^* x_b-\dot{x_a}u_b+u_a[x^t u_{t;b}+(x^t
\dot{u_t})u_b-(x^t u_t^*)x_b],
\end{equation}
where $s^{\ast{...}}_{...}=s^{...}_{...;a} x^a$ and
$A_{ab}=H^c_aH^d_b x_{c;d}$. Here $H_{ab}$ is screen projection
operator perpendicular to both $u_a$ and $x_a$, given by
\begin{equation}
H_{ab}=g_{ab}+u_au_b-x_ax_b.
\end{equation}
This has the following properties:
\begin{equation}
H^{ab}x_a=H^{ab}u_a=0,\quad\ H_c^a H_b^c=H_b^a,\quad\
H_{ab}=H_{ba},\quad\ H_a^a=2.
\end{equation}
We decompose $A_{ab}$ into its irreducible parts
\begin{eqnarray}
A_{ab}=S_{ab}+W_{ab}+\frac{1}{2}\theta^* H_{ab}\nonumber,
\end{eqnarray}
where $S_{ab}=S_{ba}$,\quad$S_a^a=0$ is the traceless part of
$A_{ab}$, $\theta^*$ is the trace of $A_{ab}$ and $W_{ab}=-W_{ba}$
is the rotation of $A_{ab}$. We can have the following relations:
\begin{equation}
S_{ab}=H_a^c H_b^d x_{(c;d)}-\frac{1}{2}\theta^* H_{ab},\quad
W_{ab}=H_a^cH_b^d x_{[c;d]},\quad \theta^*=H^{ab}x_{a;b}.
\end{equation}
The term in square parentheses, in Eq.(19), can be written as
\begin{eqnarray}
-N_b +2\omega_{tb} x^t+H_b^t \dot{x^t} \nonumber,
\end{eqnarray}
where $N_a= H_a^b(\dot{x_b}-u^*_b)$ is the Greenberg vector and
$\omega_{tb}$ is the rotational velocity. Using this value, it
follows that Eq.(19) takes the form
\begin{equation}
x_{a;b}=A_{ab}+x_a^* x_b-\dot{x_a}u_b+H_b^c
\dot{x_c}u_a+(2\omega_{tb}x^t - N^b)u_a.
\end{equation}
Also, we have
\begin{equation}
x^t u_{t;b}=2x^tu_{[t;b]}+u^*_b=-2\omega_{bt}x^t -(x_t
\dot{u^t})u_b +u_b^*.
\end{equation}
When $\xi^a=\xi x^a$, the matter inheritance symmetry, given in
Eq.(3), can be written as
\begin{equation}
T^*_{ab} + x^cT_{c(a} ln\xi_{,b)} + T_{c(a} x^c_{;b)} = 2\alpha
\xi^{-1} T_{ab}.
\end{equation}

It is known that the primary effect of a SpCKV $\xi^a=\xi x^a$ is
a well-known equation given by $ \pounds_{\xi} g_{ab} = 2 \psi
g_{ab},$  which is equivalent to the following conditions [8]
(Proof is given in Appendix B)
\begin{eqnarray}
S_{ab}&=&0,\\
x^*_a +(ln \xi)_{,a}& =& \frac {1}{2}\theta^*x_a,\\
\dot{x^a}u_a &=& \frac{-1}{2} \theta^* ,\\
N_a&=&-2\omega_{ab}x^b,\\
\psi &=& \frac{1}{2} \xi\theta^* = \xi^*.
\end{eqnarray}

\section{Timelike Matter Inheritance Vectors}

In this section, we shall prove the necessary and sufficient
conditions for the existence of timelike MIV in the model of
string cosmology and perfect fluid. In addition, we shall give the
conditions for the existence of timelike MIV which is also a
timelike CKV.

\subsection{String Cloud}
\par \noindent
{\bf Theorem:} The string cloud spacetime with energy-momentum
tensor, given by Eq.(4), admits a timelike MIV $\xi^a=\xi u^a$ if
and only if
\begin{eqnarray}
2\rho[\alpha \xi^{-1}-(ln\xi)^.]&=& \dot{\rho} ,\\
\dot{u}_c-(ln\xi)_{,c}-(ln\xi)^. u_c&=& 0,\\
2\lambda[\alpha \xi^{-1}- x_t u^{*t}] &=& \dot{\lambda},\\
2\alpha \xi^{-1}\gamma_{ab}-2\sigma_{t(a}\gamma^t_{b)}
-2\omega_{t(a}\gamma^t_{b)}\nonumber\\
-\frac{2}{3}[\theta\gamma_{ab}-h_{ab}\gamma^{cd}\sigma_{cd}
-\lambda\sigma_{ab}]&=&h^c_ah^d_b\dot{\gamma}_{cd},
\end{eqnarray}
where
\begin{equation}
\gamma^{ab} = \lambda [\frac{1}{3}h^{ab} - x^ax^b].
\end{equation}
\textbf{Proof}: First we assume that a timelike MIV exists in this
spacetime and prove that the conditions given by Eq.(31)-(34) are
satisfied.

When we substitute value of the energy-momentum tensor for string
cloud from Eq.(4) in Eq.(14) and make use of Eq.(35), we have
\begin{eqnarray}
\dot{\rho}u_au_b-\frac{\dot{\lambda}}{3}h_{ab}+2\rho[\dot{u}_{(a}u_{b)}
-u_{(a}(ln\xi)_{,b)}]&-&\frac{2\lambda}{3}[u_{(a;b)}+\dot{u}_{(a}u_{b)}]
+\dot{\gamma}_{ab}+2\gamma_{t(a}u^t_{;b)}\nonumber\\
&=&2\xi^{-1}\alpha[\rho u_a u_b -\frac{\lambda}{3}
h_{ab}+\gamma_{ab}].
\end{eqnarray}
Contracting Eq.(36) in turn with $u^a u^b,~u^a h_c^b,~h^{ab}$ and
$h^a_ch^b_d -\frac{1}{3} h^{ab}h_{cd}$, after some algebra, we
obtain
\begin{equation}
\dot{\rho}+2\rho[(ln\xi\dot{)}-\alpha \xi^{-1}]=0,
\end{equation}
\begin{equation}
h^b_c[\dot{u}_b -(ln\xi)_{,b}]=0,
\end{equation}
\begin{equation}
\dot{\lambda}+2\lambda[x_c {u^*}^c-\alpha \xi^{-1}]=0,
\end{equation}
\begin{equation}
h^a_ch^b_d\dot{\gamma}_{ab}=2\alpha \xi^{-1}
\gamma_{cd}-2\sigma_{t(c}\gamma^t_{d)}-2\omega_{t(c}\gamma^t_{d)}
-\frac{2}{3}[\theta\gamma_{cd}-h_{cd}\gamma^{ab}\sigma_{ab}
-\lambda\sigma_{cd}].
\end{equation}
\begin{description}
\item{$(i)$} Condition (31) is given by Eq.(37).
\item{$(ii)$} Expanding $h^b_c$ in Eq.(38) and multiplying by the
              rest of the expression, condition (32) turns out.
\item{$(iii)$} Condition (33) is the same as Eq.(39).
\item{$(iv)$} Eq.(40) gives condition (34).
\end{description}

Now we shall show that if the conditions (31)-(34) are satisfied,
then there must exist a timelike MIV in a string cloud spacetime.
In other words, we verify Eq.(36). When we substitute the values
from Eqs.(31)-(33) in the left hand side of Eq.(36), we obtain
\begin{equation}
2\alpha\xi^{-1}[\rho u_au_b-\frac{\lambda}{3}h_{ab}]
+\frac{2\lambda}{3}h_{ab}x_tu^{*t}-\frac{2\lambda}{3}[\sigma_{ab}
+\frac{\theta}{3}h_{ab}]+\dot{\gamma}_{ab}+2\gamma_{t(a}u^t_{;b)}
\end{equation}
If we make use of Eq.(34) in the above expression together the
value of $u_{t;a}$ from Eq.(8), we finally have the right hand
side of Eq.(36). Hence the conditions (31)-(34) are necessary and
sufficient for a string cloud spacetime to admit a timelike MIV.

The necessary and sufficient conditions for the string cloud
spacetime having a timelike MIV $\xi^a=\xi u^a$ which is also a
timelike CKV are the same as given by Eqs.(15)-(17).

\subsection{String Fluid}
\par \noindent
{\bf Theorem:} The string fluid spacetime with energy-momentum
tensor, given by Eq.(5), admits a timelike MIV $\xi^a=\xi u^a$ if
and only if
\begin{eqnarray}
(\rho_s u^a \xi)_{;a} = -\alpha T,\\
\rho_s[\dot{u}_a - (ln\xi)_{,a}-\theta u_a]=2\alpha\xi^{-1}qu_a,\\
2\alpha \xi^{-1}\gamma_{ab}-2\sigma_{t(a}\gamma^t_{b)}
-2\omega_{t(a}\gamma^t_{b)}\nonumber\\
-\frac{2}{3}[\theta\gamma_{ab}-h_{ab}\gamma^{cd}\sigma_{cd}
+(2q-\rho_s)\sigma_{ab}]&=&h^c_ah^d_b\dot{\gamma}_{cd},
\end{eqnarray}
where
\begin{equation}
\gamma^{ab} = (\rho_s +q)[\frac{1}{3}h^{ab} - x^ax^b].
\end{equation}
\textbf{Proof}: First we suppose that a string fluid spacetime
admits a timelike MIV and derive the above conditions.

When we use Eqs.(5), (14) and (45), it follows that
\begin{eqnarray}
\dot{\rho}_su_au_b+\frac{1}{3}(2\dot{q}-\dot{\rho}_s)h_{ab}
+\frac{4}{3}(q+\rho_s)\dot{u}_{(a}u_{b)}
+\dot{\gamma}_{ab}\nonumber\\
-2\rho_su_{(a}(ln\xi)_{,b)}+\frac{2}{3}(2q-\rho_s)u_{(a;b)}
+2\gamma_{c(a}u^c_{;b)}\nonumber\\
=2\alpha\xi^{-1}[\rho_s u_a u_b +
\frac{h_{ab}}{3}(2q-\rho_s)+\gamma_{ab}].
\end{eqnarray}
If we contract Eq.(46) with $u^a u^b,~u^a h_c^b,~h^{ab}$ and
$h^a_ch^b_d-\frac{1}{3} h^{ab}h_{cd}$, after some tedious
computation, we obtain
\begin{eqnarray}
\dot{\rho}_s + 2\rho_s[(ln\xi)^.-\alpha\xi^{-1}]&=&0,\\
h^b_c[\dot{u}_b-(ln\xi)_{,b}]&=&0,\\
(2\dot{q}-\dot{\rho}_s)+2\gamma^{ab}\sigma_{ab}
+\frac{2}{3}(2q-\rho_s)\theta&=&2\alpha\xi^{-1}[2q-\rho_s],\\
2\alpha \xi^{-1}\gamma_{ab}-2\sigma_{t(a}\gamma^t_{b)}
-2\omega_{t(a}\gamma^t_{b)}\nonumber\\
-\frac{2}{3}[\theta\gamma_{ab}-h_{ab}\gamma^{cd}\sigma_{cd}
+(2q-\rho_s)\sigma_{ab}]&=&h^c_a h^d_b\dot{\gamma}_{cd}.
\end{eqnarray}
\begin{description}
\item{$(i)$} Using the energy-momentum conservation law,
$T^{ab}_{;b}=0,$ in the case of string fluid, we obtain
$\dot{q}=\dot{\rho}_s$ and
$\dot{\rho}_s=-\frac{2}{3}(\rho_s+q)\theta -
\gamma^{ab}\sigma_{ab}$ which implies that
\begin{equation}
\dot{q}=-\frac{2}{3}(\rho_s+q)\theta - \gamma^{ab}\sigma_{ab}.
\end{equation}
Replacing this value of $\dot{q}$ in Eq.(49), we have
\begin{equation}
\dot{\rho}_s=-2\rho_s\theta - 2\alpha\xi^{-1}(2q-\rho_s)
\end{equation}
Comparing Eqs.(47) and (52), we have
\begin{equation}
\rho_s(ln\xi)^.=\rho_s\theta +2\alpha \xi^{-1}q
\end{equation}
which finally gives
$$(\rho_s u^a \xi)_{;a} = -\alpha T.$$
This yields the first condition given by Eq.(42).
\item{$(ii)$}
Expanding Eq.(48), we have
$$\dot{u_c}- (ln\xi)^.u_c - (ln\xi)_{,c}=0.$$
Substituting the value of $\rho_s (ln\xi)^.$ from Eq.(53) in the
above equation, we obtain the condition (43).
\item{$(iii)$} Condition (44) is the same as Eq.(50).
\end{description}

To prove that the conditions (42)-(44) are sufficient, we shall
take the left hand side of Eq.(46) given by
\begin{eqnarray}
\dot{\rho}_su_au_b+\frac{1}{3}(2\dot{q}-\dot{\rho}_s)h_{ab}
+\frac{4}{3}(q+\rho_s)\dot{u}_{(a}u_{b)}+\dot{\gamma}_{ab}\nonumber\\
-2\rho_su_{(a}(ln\xi)_{,b)}+\frac{2}{3}(2q-\rho_s)u_{(a;b)}
+2\gamma_{c(a}u^c_{;b)}.
\end{eqnarray}
Using Eq.(45) and the conditions (43)-(44), it follows that
\begin{eqnarray}
[\dot{\rho}_s +2\rho_s \theta]u^a u^b +
\frac{1}{3}(2\dot{q}-\dot{\rho}_s)h_{ab} + 4 \alpha\xi^{-1} q u_a
u_b +\nonumber\\2\alpha\xi^{-1}\gamma_{ab}
-\frac{2}{3}[(2\rho_s-q)\sigma_{ab}-h_{ab}\gamma^{cd}\sigma_{cd}]
\end{eqnarray}
Also, Eqs.(42) and (43) imply that
$$\dot{\rho}_s +2 \rho_s \theta =2\alpha\xi^{-1}(\rho_s-2q).$$
Thus the expression (55) becomes
$$2\alpha\xi^{-1}[\rho_s u_a u_b+\gamma_{ab}]+\frac{1}{3}h_{ab}
[(2\dot{\rho}_s-\dot{q})
+\frac{2}{3}(2\rho_s-q)\theta+2\gamma^{cd}\sigma_{cd}].$$ Now by
using Eq.(51), it takes the form
$$2\xi^{-1}\alpha[\rho_s u_a u_b +
\frac{h_{ab}}{3}(2q-\rho_s)+\gamma_{ab}]$$ which is the right hand
side of the Eq.(46). Hence the conditions (42)-(44) are sufficient
as well.

The string fluid spacetime admits a timelike MIV $\xi^a=\xi u^a$
which is also a timelike CKV if and only if
\begin{eqnarray}
-\frac{\rho_s\psi}{q}&=&\alpha ,\\
\sigma_{cd}&=&0,\\
\theta&=&3\psi\xi^{-1}.
\end{eqnarray}
The proof is straightforward and it follows directly by the
comparison of Eqs.(15)-(17) and Eqs.(42)-(44).

\subsection{Perfect Fluid}
\par \noindent
{\bf Theorem:} The perfect fluid spacetime with energy-momentum
tensor, given by Eq.(6), admits a timelike MIV $\xi^a=\xi u^a$ if
and only if
\begin{eqnarray}
\rho ^. &=& 2\rho[\alpha\xi^{-1} - (ln \xi)^.],\\
\dot{u}_c -(ln \xi)_{,c}-u_c(ln \xi)^.&=&0,\\
\dot{p} &=& 2p(\alpha \xi^{-1} - \frac{1}{3}\theta),\\
p\sigma_{ab} &=& 0.
\end{eqnarray}\\
\textbf{Proof}: We first assume that timelike MIV exist and prove
the above conditions. From Eqs.(6) and (14), we get
\begin{eqnarray}
(\rho+p)^.u_au_b+\dot{p}g_{ab} + 2(\rho+p)\dot{u_{(a}}u_{b)}-
2\rho u_{(a}(ln\xi)_{,b)} - 2pu_{(a;b)}\nonumber\\
=2\alpha\xi^{-1}[(\rho+p)u_a u_b+pg_{ab}].
\end{eqnarray}
Contracting Eq.(63) in turn with $u^a u^b$, $u^a h_c^b$, $h^{ab}$
and $h^a_c h^b_d - \frac{1}{3} h^{ab} h^{cd},$ we have
\begin{eqnarray}
\rho^. &=& 2\rho[\alpha \xi^{-1} - (ln \xi)^.],\\
\dot{u}_c - (ln \xi)_{,c} - (ln \xi)^. u_c&=& 0,\\
3\dot{p} + 2\theta p &=& 6\alpha \xi^{-1} p,\\
 p[(h^a_c h^b_d - \frac{1}{3} h^{ab} h^{cd})(u_{a;b} +
u_{b;a})] &=& 0.
\end{eqnarray}
\begin{description}
\item{$(i)$} Condition (59) is followed by Eq.(64)
\item{$(ii)$} Eq.(65) is the same as condition (60).
\item{$(iii)$} Condition (61) follows by Eq.(66).
\item{$(iv)$} Substituting the values of $u_{a;b}$ and
$h^a_b$ in Eq.(67), we have the condition given by Eq.(62).
\end{description}

Now we shall prove that if the conditions (59)-(62) are satisfied,
then the perfect fluid spacetime admits a MIV. Consider the left
hand side of Eq.(63) given by
$$(\rho+p)^.u_au_b+\dot{p}g_{ab} + 2(\rho+p)\dot{u_{(a}}u_{b)}-
2\rho u_{(a}(ln\xi)_{,b)} + 2pu_{(a;b)}.$$ Substituting the value
of $u_{a;b}$ and making use of Eqs.(59)-(62), we have
$$2\alpha \xi^{-1}[\rho u_a u_b+ph_{ab}].$$
Hence the conditions are necessary as well as sufficient for the
perfect fluid to admit a timelike MIV.

The perfect fluid spacetime admits a timelike MIV $\xi^a=\xi u^a$
which is also a timelike CKV if and only if the conditions given
by Eqs.(15)-(17) are satisfied.

\section{Spacelike Matter Inheritance Vectors}

This section deals with the necessary and sufficient conditions
when it admits SpMIV for string cosmology and perfect fluid.

\subsection{String Cloud}
\par \noindent
{\bf Theorem:} The string cloud spacetime with energy-momentum
tensor, admits a MIV $\xi^a=\xi x^a$ if and only if
\begin{eqnarray}
\omega_{ab}x^b &=& 0,\\
\lambda[x_a^*+(ln\xi)_{,a}] &=& \lambda x_a(ln\xi)^*,\\
\rho^*&=&-2\rho[x_a \dot{u}^a-\alpha \xi^{-1}],\\
\lambda^* &=& -2\lambda[(ln\xi)^* -\alpha \xi^{-1}].
\end{eqnarray}
\textbf{Proof}: It follows from Eqs.(4) and (25) that
\begin{eqnarray}
\rho^*u_a u_b-\lambda^*x_a x_b+2\rho_s u_{(a} u^*_{b)}- 2\lambda
x_{(a} x^*_{b)}-2\lambda x_{(a} (ln\xi)_{,b)}-2\rho x_c
u_{(a}u^c_{;b)} \nonumber\\=2\xi^{-1}\alpha[\rho u_a u_b -\lambda
x_a x_b].
\end{eqnarray}
When we make contraction of Eq.(72) with $u^a u^b$, $u^a x^b$,
$u^a H^b_c$, $x^a x^b$ and $x^aH^b_c$ in turn, the following
equations turn out
\begin{eqnarray}
\rho^*+2\rho[x_a\dot{u}^a-\alpha \xi^{-1}]&=&0,\\
\lambda[(ln\xi)^. -x_a u^{*a}]&=&0,\\
H^b_a[u^*_b -x_c u^c_{;b}]&=&0,\\
\lambda^* +2\lambda[(ln\xi)^* -\alpha \xi^{-1}]&=&0,\\
\lambda H^b_a[x^*_b+(ln\xi)_{,b}]&=&0.
\end{eqnarray}
Now we check the consistency of the necessary and sufficient
conditions given by Eqs.(68)-(71).
\begin{description}
\item{(i)} Substituting the value of $x_c u^c_{;b}$ from Eq.(24)
in Eq.(75), we obtain the condition given by Eq.(68).
\item{(ii)} Firstly, we expand Eq.(77) and then making use of Eq.(74),
we get the condition (69).
\item{(iii)} Obviously, Eq.(73) implies the condition given by Eq.(70)
directly.
\item{(iv)} Similarly, Eq.(76) follows the condition (71) directly.
\end{description}
Notice that the conditions given by Eqs.(68)-(71) satisfy Eq.(72).
Hence the conditions (68)-(71) are the necessary and sufficient
conditions for a vector $\xi^a=\xi x^a$ to be SpMIV.

It is mentioned here that a SpMIV $\xi^a=\xi x^a$ in a string
cloud spacetime is also a SpCKV if and if
\begin{eqnarray}
N_a&=&0,\\
S_{ab}&=&0,\\
\lambda(ln\xi)^* &=& \frac{\lambda \theta^*}{2},\\
x_a\dot{u}^a &=& \frac{1}{2}\theta^*,\\
\psi &=&\xi^*=\frac{1}{2}\xi \theta^*.
\end{eqnarray}
These can be easily verified by comparing Eqs.(26)-(30) and
Eqs.(68)-(71).

\subsection{String Fluid}
\par \noindent
{\bf Theorem:} The string fluid spacetime with energy-momentum
tensor has a MIV $\xi^a=\xi x^a$ if and only if
\begin{eqnarray}
\rho_s \omega_{ab}x^b &=& \frac{1}{2}q N_a,\\
q S_{ab} &=& 0,\\
x_a^*+(ln\xi)_{;a}-(x_b\dot{u}^b)x_a&=&0,\\
\rho_s\theta^* &=& -2\alpha q\xi^{-1},\\
{[\rho_s\xi x^a ]}_{;a} &=& -\alpha T.
\end{eqnarray}
{\bf Proof}: If we make use of Eqs.(5) and (25), we can write
\begin{eqnarray}
\rho_s^*[u_a u_b-x_a x_b]+q^* H_{ab}+2qx_{(a;b)}-2\rho_s x_{(a}
(ln\xi)_{;b)}\nonumber\\
+2(\rho_s +q)[u_{(a} u^*_{b)}- x_c u_{(a}
u^c_{;b}-x_{(a}x^*_{b)}]\nonumber\\
=2\xi^{-1}\alpha[(\rho_s + q)(u_a u_b -x_a x_b)+qg_{ab}].
\end{eqnarray}
Now we make contraction of Eq.(88) with $u^a u^b$, $u^a x^b$, $u^a
H^b_c$, $x^a x^b$, $x^aH^b_c$, $H^{ab}$ and $H^a_c H^b_d-
\frac{1}{2}H^{ab} H_{cd}$ in turn, the following equations are
obtained
\begin{eqnarray}
\rho^*_s +
2\rho_s[x_a\dot{u}^a-\alpha\xi^{-1}]&=&0,\\
(ln\xi)^. +x^*_a u^a&=&0,\\
qH^b_a \dot{x}_b-(\rho_s +q)H^b_a u^*_b +\rho_s H^b_a x^c
u_{c;b}&=&0,\\
\rho^*_s+2\rho_s[(ln\xi)^*
-\alpha\xi^{-1}]&=&0,\\
H^b_a[x^*_b+(ln\xi)_{,b}]&=&0,\\
q^*+q(\theta^* -2\alpha\xi^{-1})&=&0,\\
qS_{ab} &=& 0.
\end{eqnarray}
Let us now satisfy the necessary and sufficient conditions given
by Eqs.(83)-(87).
\begin{description}
\item{(i)} If we replace the value of $x^c u_{c;b}$ in Eq.(91),
we obtain Eq.(83).
\item{(ii)} It is obvious that Eq.(95) implies Eq.(84).
\item{(iii)} After expanding Eq.(93) and using Eq.(90), we arrive at
\begin{equation}
x^*_a +(ln\xi)_{,a}-(ln\xi)^* x_a=0.
\end{equation}
Now subtraction of Eq.(92) from Eq.(89) yields
\begin{equation}
(ln\xi)^* =x_a \dot{u}^a.
\end{equation}
When we use Eq.(97) in Eq.(96), we get the condition given by
Eq.(85).
\item{(iv)} The energy-momentum conservation equation for string
fluid gives the following result
\begin{eqnarray*}
q^* =-(\rho_s+q)\theta^*.
\end{eqnarray*}
When we replace this value in Eq.(94), we get Eq.(86).
\item{(v)} Since $\theta^* =H^{ab}x_{a;b}$ and hence $x_a
\dot{u}^a=x^a_{;a}-\theta^*$. When we substitute this value in
Eq.(97), we get
\begin{equation}
(ln\xi)^*=(x^a_{;a}-\theta^*).
\end{equation}
If one of the terms $\rho_s(ln\xi)^*$ in Eq.(92) is replaced by
Eq.(98), and condition (86) is used, then Eq.(92) may be written
as
\begin{eqnarray*}
(\rho_s)_{;a} \xi x^a +\rho_s(\xi_{;a} x^a + \xi
x^a_{;a})=2\alpha(\rho_s -q).
\end{eqnarray*}
Substituting $2(q-\rho_s)=T$ in the above equation, we obtain the
condition given by Eq.(87).
\end{description}
Now we explore the conditions for the string fluid when a MIV is
also a SpCKV. The string fluid spacetime admits a SpMIV $\xi^a=\xi
x^a$, which is also a SpCKV if and only if
\begin{eqnarray}
(\rho_s +q)N_a&=&0,\\
S_{ab}&=&0,\\
\xi^* &=& \psi=\frac{1}{2}\xi \theta^*,\\
\alpha &=& -\frac{\psi \rho_s}{q},\\
x_a\dot{u}^a &=& \frac{1}{2}\theta^*.
\end{eqnarray}
The proof of these results can be performed by the comparison of
Eqs.(26)-(30) and Eqs.(83)-(87). It is to be noted that in
Eq.(102), $\alpha$ turns out to be the same as in Eq.(56).

\subsection{Perfect Fluid}
\par \noindent
{\bf Theorem:} The perfect fluid spacetime with energy-momentum
tensor posesses a MIV $\xi^a=\xi x^a$ if and only if
\begin{eqnarray}
p S_{ab} &=& 0,\\
\rho \omega_{ac}x^c &=& \frac{1}{2}p N_a,\\
p[x_a^*+(ln\xi)_{,a}-(ln\xi)^*x_a] &=&0,\\
2p(ln\xi)* &=& p\theta^*,\\
\rho^* &=& 2\rho[\alpha\xi^{-1}-x_c\dot{u}^c],\\
p^*&=&p(2\alpha\xi^{-1}-\theta^* ).
\end{eqnarray}\\
\textbf{Proof}: First we assume that a perfect fluid spacetime
admits a SpMIV and show that the conditions (104)-(109) can be
obtained.

From Eqs.(6) and (25), we get
\begin{eqnarray} (\rho+p)^*u_a u_b+p^*g_{ab}
+2p[x_{(a}(ln\xi)_{,b)}+x_{(a;b)}]\nonumber\\
+2(\rho +p)[u_{(a}u_{b)}^*-x_t u_{(a} u^t_{;b)}] =2\alpha
\xi^{-1}[(\rho+p)u_a u_b+pg_{ab}].
\end{eqnarray}
Contracting Eq.(110) in turn with $u^a u^b$, $u^a x^b$, $u^a
H^b_c$, $x^a x^b$, $x^aH^b_c$, $H^{ab}$ and $H^a_c H^b_d-
\frac{1}{2}H^{ab} H_{cd}$, we have
\begin{eqnarray}
\rho^* + 2\rho[x_t\dot{u}^t-\alpha\xi^{-1}]&=&0,\\
p[(ln\xi)^. +x^*_a u^a]&=&0,\\
(\rho +p)H^b_a u^*_b -\rho H^b_a x^t u_{t;b}-pH^b_a \dot{x_b}&=&0,\\
p*+2p[(ln\xi)^* -\alpha\xi^{-1}]&=&0,\\
pH^b_a[x^*_b+(ln\xi)_{,b}]&=&0,\\
p^*+p(\theta^* -2\alpha\xi^{-1})&=&0,\\
pS_{ab} &=& 0.
\end{eqnarray}
\begin{description}
\item{$(i)$} Condition (104) is given by Eq.(117).
\item{$(ii)$} Condition (105) is derived from Eq.(113)
by substituting the value of $x^t u_{t;b}$ from Eq.(24).
\item{$(iii)$} By expanding Eq.(115) and using Eq.(112) we get
condition given by Eq.(106).
\item{$(iv)$} Subtracting Eq.(114) from Eq.(116)
gives condition (107).
\item{$(v)$} Conditions (108) and (109) are derived
directly from Eqs.(111) and (116) respectively.
\end{description}
Conversely, if the above conditions are satisfied then we show
that in the perfect fluid spacetime SpMIV exists.

Consider the left hand side of Eq.(110),
 $$(\rho+p)^*u_a u_b+p^*g_{ab}+2p[x_{(a}(ln\xi)_{,b)}+x_{(a;b)}]+2(\rho +
p)[u_{(a}u_{b)}^*-x_t u_{(a} u^t_{;b)}].$$ If we use the expansion
of $x_{a;b}$ from Eq.(23), and also use the conditions (104),
(108) and (109), we have
\begin{eqnarray}
2\alpha\xi^{-1}[(\rho+p)u_a u_b+pg_{ab}]- 2\rho u_a u_b(x_t
\dot{u}^t)-p\theta^*h_{ab}\nonumber\\
+2(\rho+p)[2u_{(a}w_{b)t}x^t +(x_t \dot{u}^t)u_a u_b]+p[\theta^*
H_{ab} -2(x_t \dot{u}^t)u_a u_b \nonumber\\-
4u_{(a}\omega_{b)t}x^t - 2u_{(a}N_{b)} + 2x_{(a}x_{b)}^*
+2x_{(a}(ln\xi)_{,b)}].\nonumber
\end{eqnarray}
Using Eqs.(105), (106) and (107), we obtain
$$2\alpha\xi^{-1}[(\rho+p)u_a u_b+pg_{ab}]$$
which implies the right hand side of Eq.(110). Hence the
conditions are necessary as well as sufficient.

The perfect fluid spacetime admits a SpMIV $\xi^a=\xi x^a$ which
is also a SpCKV if and only if the following conditions are
satisfied.
\begin{eqnarray}
S_{ab} &=& 0,\\
(ln\xi)^* &=& \frac{\theta^*}{2} ,\\
2(\rho + p)\omega_{ac}x^c&=&0,\\
x_a\dot{u^a} &=& \frac{1}{2}\theta^*,\\
\psi &=& \frac{1}{2} \xi\theta^* = \xi^*.
\end{eqnarray}
The proof follows by the comparison of Eqs.(104)-(109) and
Eqs.(26)-(30).

\section{Summary and Discussion}

This paper deals with the fundamental question of determining when
the symmetries of the geometry is inherited by all the source
terms of a prescribed matter tensor of EFEs. Physically, there is
a close connection of inheriting CKVs with the relativistic
thermodynamics of fluids since for a distribution of massless
particles in equilibrium the inverse temperature function is
inheriting CKV.

In this paper, we have found the necessary and sufficient
conditions for the existence of timelike MIVs and SPMIVs in string
cosmology and perfect fluid spacetime. In the case of timelike
MIVs, we obtain 4 conditions for the string cloud model, 3
conditions for the string fluid model and 4 conditions for the
perfect fluid model which are necessary as well as sufficient for
the existence of such vectors. In the case of SpMIVs, we obtain 5
conditions for the string cloud, 4 for the string fluid and 6 for
perfect fluid spacetime. We have also compared these conditions
with the conditions of CKV to get the conditions of MIVs which is
also a CKV. In the following we discuss these conditions in detail
for the specific cases.

\subsection{Timelike MIVs in String Cloud}

\begin{description}
\item{1.} When $\lambda=0,$ or $\rho=\rho_p,$ the case reduces
to the case of cloud of particles and the conditions are
\begin{eqnarray}
2\rho_p[\alpha \xi^{-1}-(ln\xi)^.]&=& \dot{\rho_p} ,\\
\dot{u}_c-(ln\xi)_{,c}-(ln\xi)^. u_c&=& 0.
\end{eqnarray}
\item{2.} When $\rho_p=0$ or $\rho=\lambda$, we get the conditions
for the existence of timelike MIV in geometric strings. These are
\begin{eqnarray}
2\rho[\alpha \xi^{-1}-(ln\xi)^.]&=& \dot{\rho} ,\\
\dot{u_a}-(ln\xi)_{,a}-(ln\xi)^. u_a&=& 0,\\
x_t u^{*t} &=&(ln\xi)^.,\\
2\alpha \xi^{-1}\gamma_{ab}-2\sigma_{t(a}\gamma^t_{b)}
-2\omega_{t(a}\gamma^t_{b)}\nonumber\\
-\frac{2}{3}[\theta\gamma_{ab}-h_{ab}\gamma^{cd}\sigma_{cd}
-\lambda\sigma_{ab}]&=&h^c_ah^d_b\dot{\gamma}_{cd}.
\end{eqnarray}
\end{description}

\subsection{Timelike MIVs in String Fluid}

\begin{description}
\item{1.} Eq.(42) is the kinematic equation for the string
fluid.
\item{2.} When $q=0,$ the case becomes the case of pure strings
with the following necessary and sufficient conditions to admit a
MIV
\begin{eqnarray}
(\rho_s u^a \xi)_{;a} = 2\alpha \rho_s,\\
\dot{u}_a - (ln\xi)_{a}-\theta u_a=0,\\
2\alpha \xi^{-1}\gamma_{ab}-2\sigma_{t(a}\gamma^t_{b)}
-2\omega_{t(a}\gamma^t_{b)}\nonumber\\
-\frac{2}{3}[\theta\gamma_{ab}-h_{ab}\gamma^{cd}\sigma_{cd}
-\rho_s\sigma_{ab}]&=&h^c_ah^d_b\dot{\gamma}_{cd}.
\end{eqnarray}
The conditions (129) and (130) imply that
$(ln\xi)^.=\theta=u^{*t}x_t$ and
$$2\rho_s[\alpha \xi^{-1}-(ln\xi)^.]= \dot{\rho_s}.$$ This shows
that the above conditions for string cloud and string fluid are
the same for $\lambda=\rho$ and $\rho =\rho_s.$
\end{description}

\subsection{Timelike MIVs in Perfect fluid}

\begin{description}
\item{1.} Eq.(63) gives either the shear velocity or the pressure
vanishes. When $p=0,$ it reduces to the case of dust and the
conditions become similar to the case of string cloud for
$\rho=\rho_p$. The vanshing of shear velocity implies that the
stress is zero.
\item{2.} When $\rho=p,$ it gives the stiff matter and
the corresponding conditions are
\begin{eqnarray}
\dot{p}&=& 2p[\alpha\xi^{-1} - (ln \xi)^.],\\
\dot{u}_c -(ln \xi)_{,c}&=&u_c(ln \xi)^. ,\\
\dot{p} &=& 2p(\alpha \xi^{-1} - \frac{1}{3}\theta),\\
\sigma_{ab} &=& 0.
\end{eqnarray}
We also obtain the same conditions as above for $\rho=3p$, i.e.,
the radiation case and for $\rho=-p$, i.e., the vacuum case.
\end{description}
It is to be noted that in all the cases (string cosmology and
perfect fluid), the conditions for a timelike MIV which is also a
timelike CKV are the same as given in Eqs.(15)-(17).

\subsection{ Spacelike MIVs in String Cloud}

\begin{description}
\item{1.} Eq.(68) gives
\begin{equation}
\omega_{ab}x^b =0.
\end{equation}
Since $\omega_{ab}= \eta_{abcd}\omega^cu^d$, we have by
contracting Eq.(136) with $\eta^{atef}\omega_eu_f$ that
\begin{equation}
x^a = [(\omega_bx^b)/\omega^2]\omega^a.
\end{equation}
Also, both $x^a\neq0$ and $\omega^a\neq0$, it follows that $x^a =
\pm \omega^a/\omega$ and the curves are material curves.
\item{2.} If the string tensor density is zero, i.e.,
$\lambda=0,$ then $\rho=\rho_p+\lambda$ implies that
$\rho=\rho_p.$ The case reduces to the cloud of particles and the
conditions of SpMIV given by Eqs.(68)-(71) turn out to be
\begin{eqnarray}
\omega_{ac}x^c &=& 0,\\
\rho_p^*&=&2\rho_p[\alpha \xi^{-1}-x_c \dot{u^c}].
\end{eqnarray}
When we compare the above conditions with Eqs.(15)-(17), the
conditions for a SpMIV which is also a SpCKV change into
\begin{eqnarray}
S_{ab}&=&0,\\
x_c^*+(ln\xi)_{,c}&=&\frac{1}{2}\theta^* x_c,\\
N_a&=&0,\\
x_a\dot{u^a} &=& \frac{1}{2}\theta^*,\\
\psi &=& \frac{1}{2}\xi=\xi^*.
\end{eqnarray}
\item{4.} When $\rho_p=0$, i.e., particle energy density vanishes, then
$\rho=\lambda.$ This is the case of geometric strings or Nambu
strings [18]. The conditions for the existence of SpMIV become
\begin{eqnarray}
x_c^*+(ln\xi)_{,c} &=&x_c(ln\xi)^*,\\
\omega_{ac}x^c &=& 0,\\
\rho^*&=&2\rho[\alpha \xi^{-1}-x_c \dot{u^c}],\\
x_c \dot{u^c} &=& (ln\xi)^*.
\end{eqnarray}
Also, the conditions for a SpMIV which is also a SpCKV takes the
form
\begin{eqnarray}
S_{ab}&=&0,\\
N_a&=&0,\\
(ln\xi)^* &=& \frac{\theta^*}{2}=x_c \dot{u^c},\\
\psi &=&\frac{1}{2}\xi\theta^*= \xi^*.
\end{eqnarray}
\end{description}

\subsection{ Spacelike MIVs in String Fluid}

\begin{description}
\item{1.} It follows from Eq.(84) that either $q=0$ or $S_{ab}
=0$. If $q=0$, the case reduces to the case of pure strings and
$S_{ab} =0$ implies that the shear of a spacelike congruence,
generated by a MIV, vanishes.

\item{2.} If we take $\omega=0$, Eq.(83) implies that $q N_a =0$ which
gives two possibilities either  $q=0$ or $N_a =0$. The case $q=0$
is the same as above and $N_a =0$ implies the integral curve,
i.e., $x^a$ are material curves and the string fluid forms the
two-surface.

\item{3.} When we assume that $\omega\neq0$ and $N_a =0$, we get $x^a =
\pm \omega^a/\omega$ and curves are material curves.
\end{description}

\subsection{ Spacelike MIVs in Perfect Fluid}

\begin{description}
\item{1.} If $N_a=0$, i.e., the integral curves are material curves,
then the condition (105) gives $\rho\omega_{ac}x^c = 0$ which
implies that either $\rho=0$ or $\omega_{ac}x^c=0.$ If
$\omega_{ac}x^c=0,$ we obtain the similar results as for the case
of string cloud.
\item{2.} When $p=0,$ the case reduces to the case of dust. In this case
we get the same necessary and sufficient conditions for the
existence of the MIV as in the case of cloud of particles with
$\rho_p=\rho.$
\item{3.} The case $\rho=p$ in a perfect fluid implies stiff
matter. The necessary and sufficient conditions of the existence
of MIV in this case are
\begin{eqnarray}
S_{ab} &=& 0,\\
\omega_{ac}x^c &=& \frac{1}{2}N_a,\\
x_a^*+(ln\xi)_{,a}-(ln\xi)^*x_a &=&0,\\
2(ln\xi)* &=& \theta^*,\\
p^* &=& 2p[\alpha\xi^{-1}-x_c\dot{u}^c],\\
p^*&=&p(2\alpha\xi^{-1}-\theta^* ).
\end{eqnarray}
\item{4.} $\rho=3p$ implies the case of radiation and the
conditions of the existence of MIV reduce to
\begin{eqnarray}
S_{ab} &=& 0,\\
\omega_{ac}x^c &=& \frac{1}{6}N_a,\\
x_a^*+(ln\xi)_{,a}-(ln\xi)^*x_a &=&0,\\
2(ln\xi)* &=&\theta^*,\\
p^* &=& 2p[\alpha\xi^{-1}-x_c\dot{u}^c],\\
p^*&=&p(2\alpha\xi^{-1}-\theta^* ).
\end{eqnarray}
\item{5.} $\rho=-p$ in a perfect fluid gives vacuum state. The
necessary and sufficient conditions of the existence of MIV in
this case are
\begin{eqnarray}
S_{ab} &=& 0,\\
N_a+ 2\omega_{ac}x^c&=& 0,\\
x_a^*+(ln\xi)_{,a}-(ln\xi)^*x_a &=&0,\\
2(ln\xi)* &=&\theta^*,\\
p^* &=& 2p[\alpha\xi^{-1}-x_c\dot{u}^c],\\
p^*&=&p(2\alpha\xi^{-1}-\theta^* ).
\end{eqnarray}
The necessary and sufficient conditions for the existence of MIV
which is also a CKV are the same in the cases (3), (4) and (5)
given as
\begin{eqnarray}
S_{ab} &=& 0,\\
(ln\xi)^* &=& \frac{\theta^*}{2} ,\\
\omega_{ac}x^c&=&0,\\
x_a\dot{u^a} &=& \frac{1}{2}\theta^*,\\
\psi &=& \frac{1}{2} \xi\theta^* = \xi^*.
\end{eqnarray}
\end{description}
It is mentioned here that for $\alpha=0$, we obtain the conditions
of matter collineations in each case of spacelike and timelike
MIVs for the models of string cloud, string fluid and perfect
fluid.

We have obtained the conditions for the existence of MIVs in the
models of string cloud, string fluid and perfect fluid. These
conditions can be used as restriction for the EFEs. Since the
non-linearity of EFEs ceases to extract their exact solution, the
restricted equations may give interesting solution in respective
spacetimes. Matter inheritance symmetry for null fluid spacetimes
can be defined. Cylindrically symmetric and spherically symmetric
spacetimes can be classified by the MIVs. It would be worth
interesting to look for the necessary and sufficient conditions
for the existence of null MIVs in different cosmological models.

\renewcommand{\theequation}{A\arabic{equation}}
\setcounter{equation}{0}

\section*{Appendix A}

\textbf{Conditions for Timelike CKVs}\\Here we prove the necessary
and sufficient conditions for a timelike CKV.\\
\par \noindent
{\bf Theorem:} The primary effect of a timelike CKV $\xi^a=\xi
u^a$ is a well-known equation $\pounds_{\xi} g_{ab} = 2\psi
g_{ab},$ which is equivalent to the following conditions [21]
\begin{eqnarray}
\dot{u}_a -(ln \xi)_{;a}&=& u_a(ln\xi)^.=\xi^{-1}\psi u_a,\\
\theta&=& 3\xi^{-1}\psi,\\
\sigma_{cd} &=&0.
\end{eqnarray}
\textbf{Proof}: The symmetry equation of CKV is $$\pounds_{\xi}
g_{ab} = 2\psi g_{ab}$$ which implies that
$$\xi_{a;b}+\xi_{b;a}=2\psi g_{ab}.$$
Contracting the above equation in turn with $u^a u^b$,$u^a h_c^b$,
$h^{ab}$ and $ h^a_ch^b_d -\frac{1}{3} h^{ab}h_{cd}$ and applying
$\xi^a=\xi u^a$, we have
\begin{eqnarray}
u_a(ln\xi)^.&=&\xi^{-1}\psi,\\
\dot{u}_a -(ln \xi)_{;a}-(ln \xi)^. u_a &=&0,\\
\theta&=& 3\xi^{-1}\psi,\\
\sigma_{cd} &=&0.
\end{eqnarray}
\begin{description}
\item ($i$) Making use of (A5) in (A4), we have the condition (A1).
\item ($ii$) Eqs.(A6) and (A7) directly imply the conditions (A2) and
(A3).
\end{description}
By putting these conditions in $\xi_{a;b}+\xi_{b;a}$, where
$\xi^a=\xi u^a,$ we obtain $2 \psi g_{ab}$ and so the conditions
are necessary as well as sufficient.

When $\psi=0,$ the conditions (A1)-(A3) reduces to the necessary
and sufficient condition for the existence of timelike KV.

\renewcommand{\theequation}{B\arabic{equation}}
\setcounter{equation}{0}
\section*{Appendix B}

\textbf{Conditions for Spacelike CKVs}\\Here we prove the
necessary and sufficient conditions for a SpCKV.\\
\par \noindent
{\bf Theorem:} The primary effect of a SpCKV $\xi^a=\xi x^a$ is a
well-known equation $\pounds_{\xi} g_{ab} = 2\psi g_{ab},$ which
is equivalent to the following conditions [8]
\begin{eqnarray}
S_{ab}&=&0,\\
x^*_a +(ln \xi)_{,a}& =& \frac {1}{2}\theta^*x_a,\\
u_a \dot{x^a} &=& -\frac{1}{2} \theta^*,\\
N_a&=&-2\omega_{ab}x^b,\\
\psi &=& \frac{1}{2} \xi\theta^* = \xi^*.
\end{eqnarray}
\textbf{Proof}: The symmetry equation of CKV is $$\pounds_{\xi}
g_{ab} = 2\psi g_{ab}$$ which implies that
$$\xi_{a;b}+\xi_{b;a}=2\psi g_{ab}.$$
Contracting the above equation in turn with $u^a u^b$, $u^a x^b$,
$u^a H^{bc}$, $x^a x^b$, $x^aH^{bc}$ and $H^{ac}H^{bd}$ and
applying $\xi^a=\xi x^a$, we have
\begin{eqnarray}
u_a \dot{x^a} &=& -\frac{\psi}{\xi},\\
u^a[x^*_a + (ln\xi)_{,a}]&=&0,\\
H^{ab}[\dot{x}_b +u^t x_{t;b}]&=&0,\\
\xi^* &=&\psi,\\
H^{ab}[x^*_b + (ln\xi)_{,b}]&=&0,\\
S_{ab}+\frac{1}{2}[\theta^*-\frac{2\psi}{\xi}]H_{ab}&=&0.
\end{eqnarray}
\begin{description}
\item ($i$) Multiply (B11) by $g^{ab}$, we have
$$S^a_a+\frac{1}{2}[\theta^* - \frac{2\psi}{\xi}]H^a_a=0.$$ This
implies that $\theta^*=\frac{2\psi}{\xi}.$ This equation with
Eq.(B9) gives condition (B5).
\item ($ii$) Replacing the value of $\psi$ from Eq.(B5) in
Eq.(B6), we have condition (B3).
\item ($iii$) Eqs.(B7) and (B10) implies that
$$x^*_a + (ln\xi)_{;a}-x_a(ln\xi)^*=0.$$
As $$(ln\xi)^*=\frac{\xi^*}{\xi}=\frac{\theta^*}{2},$$ the above
equation gives $$x^*_a + (ln\xi)_{,a}-\frac{\theta^*x_a}{2}=0$$ or
$$x^*_a +(ln \xi)_{,a}= \frac {1}{2}\theta^*x_a$$ which is
condition (B2).
\item ($iv$) Substituting the value of $\psi$ from Eq.(B5) in Eq.(B11),
Eq.(B1) turns out.
\item ($v$) On expanding Eq.(B8), we have $$H^b_a(\dot{x_b}+u^*_b)
+2\omega_{ab}x^b=0$$
which gives condition (B4).
\end{description}
Substituting these conditions in $\xi_{a;b}+\xi_{b;a}$ where
$\xi^a=\xi x^a,$ we obtain $2 \psi g_{ab}$ and so the conditions
are necessary as well as sufficient. When $\psi=0$, the conditions
(B1)-(B5) reduce to the necessary and sufficient conditions of
SpKVs.

\newpage
{\bf \large References}

\begin{description}

\item{[1]} Duggal, K.L.: J. Math. Phys. \textbf{33}(1992)2989.

\item{[2]} Herrera, L. and Ponce de, L.J.: J. Math. Phys. \textbf{26}(1985)778;
\textbf{26}(1985)2018; \textbf{26}(1985)2847.

\item{[3]} Maartens, R., Mason, D.P. and Tsamparlis, M.: J. Math. Phys.
\textbf{27}(1986)2987.

\item{[4]} Coley, A.A. and Tupper, B.O.J.: J. Math. Phys. \textbf{30}(1989)2616.

\item{[5]} Carot, J., Coley, A.A., and Sintes, A.: Gen. Rel. Grav.
\textbf{28}(1986)311.

\item{[6]} Duggal, K.L.: Acta Appl. Math. \textbf{31}(1993)225.

\item{[7]} Greenberg, P.J.: J. Math. Anal. Appl. \textbf{30}(1970)128.

\item{[8]} Mason, D.P. and Tsamparlis, M.: J. Math. Phys. \textbf{24}(1983)1577.

\item{[9]} Yavuz, $\dot{I}$. and Yilmaz, $\dot{I}$.: Gen. Rel. Grav.
\textbf{29}(1997)1295.

\item{[10]} Yilmaz, $\dot{I}$., Tarhan, $\dot{I}$., Yavuz, $\dot{I}$.,
Baysal, H. and Camci, U.: Int. J. Mod. Phys. D\textbf{8}(1999)659.

\item{[11]} Baysal, H., Camci, U.,  Yilmaz, $\dot{I}$. and  Tarhan, $\dot{I}$.:
 Int. J. Mod. Phys. D\textbf{11}(2002)463.

\item{[12]} Yilmaz, $\dot{I}$.: Int. J. Mod. Phys. \textbf{D10}(2001)681.

\item{[13]} Baysal, H. and Yilmaz, $\dot{I}$: Class. Quantum Grav. \textbf{19}
(2002)6435.

\item{[14]} Baysal, H. and Yilmaz, $\dot{I}$: Turk J. Phys \textbf{27}(2003)83.

\item{[15]} Zeldovich, Ya B.: Mon. Not. R. Astron. Soc. \textbf{192}(1980)663.

\item{[16]} Kibble, T.W.S.: J. Phys. A: Math. Gen. \textbf{9}(1976)1387.

\item{[17]} Letelier, Patricio S.: Phys. Rev. \textbf{D22}(1980)807.

\item{[18]} Letelier, Patricio S.: Phys. Rev. \textbf{D28}(1983)2414.

\item{[19]} Letelier, Patricio S.: Nuovo Cimento \textbf{63B}(1981)519.

\item{[20]} Stephani, Hans: {\it General Relativity: An Introduction
to the Theory of the Gravitational Field} (Cambridge University
Press, 1990).

\item{[21]} Oliver, D.R. and Davis, W.R.: Gen. Rel. Grav. \textbf{8}(1977)905.

\item{[22]} Duggal, K.L. and Sharma, R.: {\it Symmetries of Spacetimes
and Riemannian Manifolds} (Kluwer Academic Publishers, 1999).

\end{description}

\end{document}